# Experimental Test on Edwards Volume Ensemble of Tapped Granular Packings

Ye Yuan,[1] Yi Xing,[1] Jie Zheng,[1] Zhifeng Li,[1] Houfei Yuan,[1] Shuyang Zhang,[1] Zhikun Zeng,[1] Chengjie Xia,[2] Hua Tong,[1,3] Walter Kob,[1,4] Jie Zhang,[1,5] Yujie Wang[1,6*]

[1]*School of Physics and Astronomy, Shanghai Jiao Tong University, Shanghai 200240, China*
[2]*School of Physics and Electronic Science, East China Normal University, Shanghai 200241, China*
[3]*Department of Physics, University of Science and Technology of China, Hefei 230026, China*
[4]*Laboratoire Charles Coulomb, UMR 5521, University of Montpellier and CNRS, 34095 Montpellier, France*
[5]*Institute of Natural Sciences, Shanghai Jiao Tong University, Shanghai 200240, China*
[6]*Materials Genome Initiative Center, Shanghai Jiao Tong University, Shanghai 200240, China*

Using X-ray tomography, we experimentally investigate granular packings subject to mechanical tapping for three types of beads with different friction coefficients. We validate Edwards volume ensemble in these three-dimensional granular systems and establish a granular version of thermodynamic zeroth law. Within Edwards framework, we also explicitly clarify how friction influences granular statistical mechanics as modifying the density of states, which allows us to determine the entropy as a function of packing fraction and friction subsequently. Additionally, we obtain a granular jamming phase diagram based on geometric coordination number and packing fraction.

Granular packings are intrinsically non-equilibrium, as they stay in mechanically stable configurations in absence of external driving. However, when a granular system is subjected to a specific excitation protocol like consecutive tapping, it can reach a stationary state whose packing fraction fluctuates around a specific value independent of the preparation history [1-5], reminiscent of a system in thermal equilibrium [6,7]. Edwards and co-workers proposed a statistical mechanics approach to account for this observation, postulating that for granular systems volume is a conserved quantity, analogous to energy for systems in thermal equilibrium [7-9]. In this approach, the corresponding conjugate to volume is a temperature-like variable called "compactivity", which describes the compaction capability for a granular system. This framework was later extended to include contact forces in a jammed configuration which introduces the force or stress ensemble [10,11], although its relationship with the volume ensemble remains debated [12-14].

The Edwards ensemble approach has been investigated by a number of experimental and numerical studies. It is found that when a granular assembly reaches the steady state, local quantities like volume [15-21] or stress [11,20,22] possess a Boltzmann-like distribution. Furthermore, a numerical study has shown that mechanically stable frictionless packings are equally probable at the jamming point, verifying Edwards' original conjecture that the logarithm of the number of mechanical stable packings gives granular entropy [23]. Other simulations have also found that the Edwards compactivity equals the dynamic effective temperature in slowly driven granular materials, leading to a granular version of the fluctuation-dissipation theorem [24,25]. This justification is critical since it proves the usefulness of thermodynamic framework for the development of a constitutive theory for granular materials [26,27].

Nonetheless, a systematic experimental verification of Edwards volume ensemble in three dimensions (3D) is still lacking. It remains controversial whether compactivity defined over the volume ensemble satisfies the thermodynamic zeroth law [20]. Furthermore, previous experimental investigations mainly focused on a single species of granular beads, i.e., with the same friction coefficient. The explicit role friction plays on granular statistical mechanics remains little understood [28].

In this Letter, we address these unresolved issues by analysing experimentally tapped granular packings in 3D using X-ray tomography technique [5,15,29,30]. We employ three types of monodisperse beads with different friction coefficients. Our results support the validity of Edwards volume ensemble and the existence of thermodynamic zeroth law. Furthermore, we find that at a given volume, both the density of states and the entropy increase with friction, which elucidates the specific way that friction influences granular statistical mechanics. Additionally, we obtain a granular jamming phase diagram based on geometric coordination number and packing fraction, which sheds light on the potential coupling between the volume and stress ensembles.

We generate compacted disordered granular packings via mechanical tapping for three types of monodisperse spherical beads in a dry cylindrical container. We employ the Acrylonitrile Butadiene Styrene plastic (ABS) beads, the 3D-printed plastic (3DP) beads (ProJet MJP 2500 Plus), and the 3D-printed plastic beads with bumpy surface (BUMP). The BUMP particle is designed by decorating a sphere of diameter $d$ with 100 small spheres of diameter $0.1d$ on its surface (see Supplemental Materials [31]). We use the BUMP bead to mimic a particle with large surface friction while maintaining a geometrical shape almost identical to an ideal sphere. The bead diameters are $d = 6$ mm for 3DP beads, $d = 5$ mm for ABS and BUMP beads respectively. Their effective friction coefficients follow $\mu_{\text{BUMP}} > \mu_{\text{3DP}} > \mu_{\text{ABS}}$ as measured by repose angle measurements [31]. The inner diameter of the cylindrical container is $D = 140$ mm and the height of the packing is roughly 140 ~ 210 mm. We glue ABS semi-spheres with $d = 5$ mm and 8 mm on the bottom and side walls of the container to avoid crystallization of the packing structure during the whole compaction process.

A mechanical shaker is used to excite the granular packings, with tap intensity $\Gamma = 2g$ ~ $16g$, where $g$ is the gravitational acceleration constant. A tap cycle consists of a 200 ms pulse and a 1.5 s interval to fully settle the system. The compaction process starts from an initial poured-in random packing structure. Depending on $\Gamma$, the tap numbers needed to reach the stationary states are about 10 ~ 4000. In general, a stronger tap intensity leads to a looser stationary-state packing. X-ray tomography scans are performed on the stationary-state packings using a medical CT scanner (UEG Medical Group Co. Ltd., 0.2 mm spatial resolution). Subsequently, 3D packing structures are obtained following similar image processing procedures as previous studies [5,32]. For the analysis presented here, each packing consists of 3000 ~ 6000 particles, after excluding particles within $2.5d$ from the container boundary and free surface. Statistics for each stationary state are the average result over 10 ~ 30 independent realizations.

The global packing fraction $\phi$ is calculated using $\phi = \sum v_p / \sum v_{voro}$, where $v_p$ and $v_{voro}$ are the particle volume and its associated Voronoi cell volume. For simplicity, in the following we set $v_p$ as unity. As shown in Fig. 1(a), the relation between $\Gamma$ and $\phi$ is not universal, i.e., identical $\Gamma$ would generate different $\phi$ for all three systems. Specifically, $\phi = 0.605 \sim 0.640$ for ABS, $\phi = 0.587 \sim 0.640$ for 3DP, and $\phi = 0.565 \sim 0.625$ for BUMP. The associated lower bounds $\phi^* = 0.605$, 0.587, and 0.565 clearly depend on the friction coefficient $\mu$ for different beads. We verify that $\phi^*$ can be independently reproduced by a hopper deposition protocol, which is another typical way to generate very loose packings [31]. The quantitative agreement between two protocols on values of $\phi^*$ indicates that $\phi^*$ indeed marks the onset of mechanical stability for packings under gravity. The $\phi$ range obtained fully covers the range bounded between $\phi_{\text{RLP}} \approx 0.56$ for the random loose packing (RLP) and $\phi_{\text{RCP}} \approx 0.64$ for the random close packing (RCP) [5,30,33-36]. The sufficient overlaps of the $\phi$ range between different systems can allow us to investigate the role friction plays systematically.

If Edwards volume ensemble is valid in our system, compactivity should be an intensive quantity, whose value and the associated volume fluctuation should be independent of the system size. However, since Voronoi volumes of neighboring particles are correlated, single-particle statistics cannot satisfy this requirement [21,37]. To avoid this issue, we calculate the total Voronoi volume $V$ of particles within a coarse-grain spherical region of diameter $d(m/\phi)^{1/3}$ around each particle, where $m$ is the number of particles within. In practice, we adopt $m = 15$, which sufficiently suppresses the Voronoi volume correlations [31].

According to Edwards volume ensemble, the probability of finding a system volume $V$ is given by a Boltzmann-like distribution

$$P_\mu(V) = \frac{\Omega_\mu(V)}{Z_\mu(\chi)} e^{-V/\chi}, \tag{1}$$

where $\chi$ is the compactivity, $\Omega_\mu(V)$ is the density of states, $Z_\mu(\chi)$ is the partition function, and the subscript $\mu$ indicates the type of bead. In Fig. 1(b), we show the probability distribution function $P_\mu(V)$ for four representative packings. All these distributions can be well fitted by a $k$-Gamma function as previously reported [16], whose exponential tails are consistent with the Boltzmann-like distribution of Eq. (1). We note that $P_\mu(V)$ of ABS and BUMP at $\phi = 0.605$ are very similar to each other, indicating that $P_\mu(V)$ mainly depends on $\phi$. Nevertheless, since $\Omega_\mu(V)$ depends on the type of bead considered, this similarity does not necessarily lead to a simple relation between $\chi$ and $\phi$. With increasing $m$, the distribution gradually becomes Gaussian-like which might complicate the justification of the Boltzmann-like distribution [16,18]. Fortunately it is not an issue in our case [31].

To measure $\chi$, we use the overlapping histogram method which considers the ratio between $P_\mu(V)$ and $P_\mu^r(V)$ of a reference state with the same $\mu$ [9],

$$\frac{P_\mu^r(V)}{P_\mu(V)} = \frac{Z_\mu(\chi)}{Z_\mu(\chi^r)} e^{(\frac{1}{\chi} - \frac{1}{\chi^r})V}, \qquad (2)$$

where $\chi^r$ is the compactivity of the reference state. In Fig. 1(c), the logarithms of the left hand side of Eq. (2) is plotted as a function of $V/m$ for the configurations shown in Fig. 1(b) and the clear presence of linear regimes is recognized with the slopes corresponding to $\left(\frac{1}{\chi} - \frac{1}{\chi^r}\right)m$. Hence, this allows us to determine $\chi$ if the reference state $P_\mu^r(V)$ and its $\chi^r$ can be identified. In analogy to the relation between the energy fluctuation and the specific heat in thermal equilibrium systems [5,9], $\chi$ can also be determined from the intensive volume fluctuation $var(V) = \sigma_V^2/m$, where $\sigma_V^2$ is the variance of $V$. The associated relation is

$$\frac{1}{\chi(\phi)} - \frac{1}{\chi^r} = \int_{\phi^r}^{\phi} \frac{d\varphi}{\varphi^2 var(V)}, \qquad (3)$$

where $\phi^r$ is the packing fraction of the reference state. As shown in Fig. 1(d), surprisingly, $var(V)$ for all three systems nearly collapse on a master curve, despite their rather different values of $\mu$, in agreement with the $P_\mu(V)$ dependency on $\phi$ from Fig. 1(b). This master curve can be described well by a cubic polynomial fit given by $var(V) = 9.718 - 45.335\phi + 70.959\phi^2 - 37.225\phi^3$ (solid curve in Fig. 1(d)). In previous work, the reference state (i.e., infinite $\chi$) used in Eqs. (2) and (3) normally has been chosen to be the RLP state with $\phi_{\text{RLP}} \approx 0.56$ [9]. Instead, we believe it is more suitable to define the infinite $\chi$ reference states as $\phi^r = \phi^* = 0.565, 0.587$, and $0.605$ for BUMP, 3DP, and ABS, respectively, i.e., the loosest packing that can be reached by certain bead depends on $\mu$ [17]. It turns out $\chi^{-1}$ calculated via Eq. (2) (symbols) are consistent with those via Eq. (3) (solid curves) for all three systems, as shown in Fig. 2(a), which demonstrates the equivalency of two methods [20,21]. The relationship between $\chi$ and $\phi$ for different systems (inset of Fig. 2(a)) qualitatively agrees with previous numerical studies [17,35]. The nice agreement of the probability distribution of *V* with the function form of Eq. (1) and the consistency of results of two independent protocols to calculate $\chi$ therefore strongly support the validity of Edwards volume ensemble in our system.

In Fig. 2(b), we observe nice one-to-one correspondence between $\chi$ and tap intensity $\Gamma$ for all three particle species within experimental uncertainty, where the margins of error are calculated assuming a 0.01 uncertainty for values of $\phi^*$ in all three systems [31]. This result implies that two different granular packings (i.e., with different $\mu$) under the same excitation intensity, i.e., same external heat bath, would reach the similar "temperature" (i.e. $\chi$). Thus, granular version of zeroth law of thermodynamics is fulfilled in our systems. It was previously reported that two subsystems consisting of different frictional disks in a single packing reach different $\chi$ under quasistatic compression, with almost one-fold difference in $\chi$ over the whole $\phi$ range investigated [20]. This apparent inconsistency might originate from the fact that quasistatic compression significantly prohibits structural rearrangement, which hampers two subsystems to reach thermal equilibrium under the volume ensemble. There also exists the possibility that discrepancy originates from different definitions for the reference state for infinite $\chi$ in two studies. If $\chi_0$ is calculated by Eq. (3) using the constant RLP reference with $\phi^r = 0.565$ for all three systems, it simply becomes a function of $\phi$, since $var(V)$ only depends on $\phi$ as shown in Fig. 1(d). In this case, the one-to-one correspondence between $\Gamma$ and compactivity is lost (inset of Fig. 2(b)). The emergence of thermodynamic zeroth law for granular materials is quite striking given that the dissipative dynamics and microscopic structures of different systems under the same $\Gamma$ are rather distinctive as indicated by the relation between $\chi$ and $\phi$ in the inset of Fig. 2(a).

Until now, we have not yet discussed the details how friction influences the statistical behavior of our system. According to Eq. (1), since $\Omega_\mu(V) = P_\mu(V)e^{V/\chi}Z_\mu(\chi)$, the *V*-dependency of $\Omega_\mu(V)$ is given by the $P_\mu(V)e^{V/\chi}$ part only. So when plotting $P_\mu(V)e^{V/\chi}$ as a function of *V* for a given $\mu$ with different $\chi$, they should collapse onto a master curve after being scaled by a certain factor. Figure 3(a) shows that this is indeed the case, which suggests that the functional form of Eq. (1) is valid. Additionally, since the scaling factor of $P_\mu(V)e^{V/\chi}$ at finite $\chi$ needed to collapse onto its $\chi \to \infty$ counterpart is the $\chi$-dependent $Z_\mu(\chi)/Z_\mu(\infty)$, we can then obtain $Z_\mu(\chi)$ up to a scaling constant $Z_\mu(\infty)$.

Similarly, $\Omega_\mu(V)$ is the product of the normalized master probability distribution $P_\mu^{\chi\to\infty}(V)$ with $Z_\mu(\infty)$ since at infinite $\chi$, $\Omega_\mu(V) = P_\mu^{\chi\to\infty}(V)Z_\mu(\infty)$. We can also obtain $\Omega_\mu(V)$ for all three systems up to the scaling constant $Z_\mu(\infty)$. A simple inspection of the master curve distributions for the three systems clearly shows that the density of states $\Omega_\mu(V)$ depends on $\mu$.

Once we obtain $Z_\mu(\chi)/Z_\mu(\infty)$, we can then evaluate the free energy by $F = -\chi \ln(Z_\mu(\chi))$ with an additive term $\chi\ln(Z_\mu(\infty))$. In Fig. 3(b), we show $F$ as a function of $\chi^{-1}$ and this dependence can be well fitted by the functional form $y = c_1 + c_2 x^{c_3}$. From $F$ we can now obtain the entropy by $S = -\partial F/\partial \chi$ and also $S(\chi)$ will have an unknown additive constant $\ln(Z_\mu(\infty))$. To determine this constant, we postulate that $S_{RCP}$ is zero for all three systems (It was proposed that $S_{RCP}$ is a finite constant [17]. However, its value does not qualitatively affect the discussion here.), from which $Z_\mu(\infty)$ can also be determined. The resulting $S(\phi)$ are shown in Fig. 3(c) (symbols), in qualitative agreements with previous numerical simulations [17]. Complemental to this approach, entropy can also be measured using the fluctuation method [5]:

$$S(\phi) - S_{RCP} = \int_\phi^{\phi_{RCP}} \frac{d\varphi}{\varphi^2 \chi(\varphi)}, \qquad (4)$$

as indicated by the dashed curves in Fig. 3(c). The agreement of both approaches gives additional strong support for using Edwards statistical mechanics description of our system. For a given $\phi$, the value of $S(\phi)$ increases with increasing $\mu$, which is reasonable since a larger $\mu$ can stabilize more packings at a given $\phi$. This can also be confirmed by explicitly calculating the density of states $\Omega_\mu(V)$ which can be obtained by multiplying the curves in Fig. 3(a) by the constant $Z_\mu(\infty)$, as shown in Fig. 3(d). This result demonstrates that friction strongly influences $\Omega_\mu(V)$ as that a large $\mu$ allows for the existence of more mechanically stable states at a given $V$. This result agrees qualitatively with a recent experiment on 2D disk packings [28].

Based on Edwards volume ensemble, a jamming phase diagram for frictional granular materials can be constructed by a mean-field theory [35]. To construct a similar phase diagram, we determine the geometric coordination number $z$ for all the experimental packings. The procedure follows a strictly identical criterion for all packings to exclude the errors introduced by image processing [27]. In Fig. 4, we show the relation between $z$ and $\phi$ for all three systems, together with our simulation results of packings at jamming onset with varying $\mu$ [31]. Notable differences exist between previous study and our results. Specifically, in our study, $z$ is no longer simply a function of $\phi$ as proposed in the mean-field theory [35], but shows complex behaviors. We observe an onset (loosest) packing state with $z^*(\phi^*)$ for each bead. $z^* = 4.2$ for the loosest BUMP packings with $\phi^* = 0.565$, which is close to the isostatic $z = 4$ for the jamming transition as $\mu \to \infty$, i.e., RLP. For the other two systems we find $z^*$ at larger values at their respective loosest packings. Interestingly, $z^*(\phi^*)$ is rather close to our simulation results for frictional jamming onset, indicating a unique relation between $z^*$, $\phi^*$, and $\mu$. Beyond this onset state, for any of the system, $z$ grows with $\phi$ in a similar manner between the jamming onset and RCP. Introducing the scaled variables $\tilde{z} = (z - z^*)/(z^0 - z^*)$ and $\tilde{\phi} = (\phi - \phi^*)/(\phi_{RCP} - \phi^*)$, where $z^0 = 6$ is the isostatic coordination number for RCP, we find that all three systems satisfy an empirical relation $\tilde{z} = \tilde{\phi}^{2.5}$, as shown in the inset of Fig. 4. For a given $\phi$, the upper bound of $z$ is formed by the corresponding onset $z^*(\phi^*)$, which reflects the geometrical constraint for a disordered hard-sphere configuration. At the same $\phi$, systems with different $\mu$ will have different $z$, i.e., $z(\phi, \mu_1) < z(\phi, \mu_2)$ if $\mu_1 > \mu_2$, which indicates that packings with a larger $\mu$ possess more mechanically stable configurations with small $z$. This result is rather different from previous study in which $z$ only depends on $\phi$ and mechanically stable configurations are instead related with mechanical coordination number [35]. Our findings clearly suggest that $z$ already encodes information about mechanical stability, which implies the coupling between the stress and volume ensembles for $\phi < \phi_{RCP}$ [12].

Also included in Fig. 4 are iso-compactivity lines (dashed curves) with $\chi = 0.8, 0.4$, and $0.2$. They are analogous with the ones in the phase diagram with mechanical coordination number as the vertical axis (Fig. 1 in [35]). This is not surprising for the same reason just mentioned. Note that a unique RCP state with $z^0 = 6$, $\phi \approx 0.64$ is approached as $\chi \to 0$, irrespective of $\mu$. Such convergence is further complemented by the behaviors of other structural parameters [31]. Therefore, geometric packing structure and its fluctuations at RCP are identical for systems with different $\mu$, reinforcing previous postulate that entropy $S_{RCP}$ is constant at RCP for different systems, and stress and volume ensembles are decoupled at RCP [9].

In summary, by systematically studying tapped granular packings, we validate Edwards volume ensemble in 3D granular systems and establish a granular version of thermodynamic zeroth law. Additionally, we clarify how friction influences granular statistical mechanics as modifying the density of states. It subsequently allows us to determine the entropy as a function of packing fraction and friction, which can help us establish the connections between microscopic and macroscopic properties of granular materials. We believe these results are quite important since it shows that Edwards ensemble is indeed a coherent framework to describe the statistical properties of granular systems and it will help any future development of constitutive law of granular materials based on thermodynamic framework.

The work is supported by the National Natural Science Foundation of China (No. 11974240) and Shanghai Science and Technology Commission (No. 19XD1402100).

\* Corresponding author

yujiewang@sjtu.edu.cn

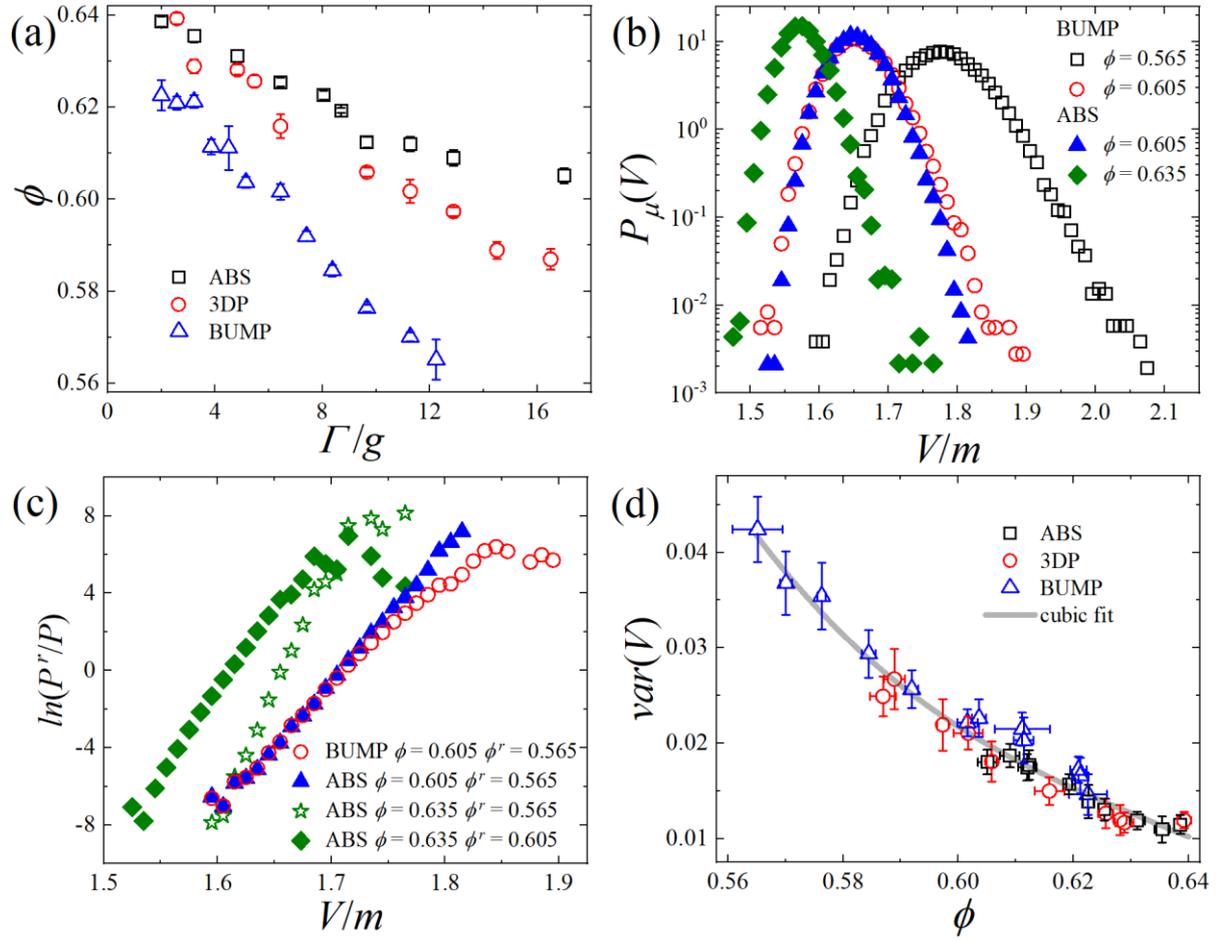

FIG. 1. (a) Packing fraction $\phi$ as a function of tap intensity $\Gamma$ for the three systems (ABS, 3DP and BUMP). (b) Probability distribution functions $P_\mu(V)$ for four representative packings. (c) Relationship between $\ln(\frac{P^r_\mu(V)}{P_\mu(V)})$ and $V/m$ for packings in (b). (d) Volume fluctuation $var(V)$ as a function of $\phi$ (symbols). Solid curve is the cubic polynomial fit (see main text).

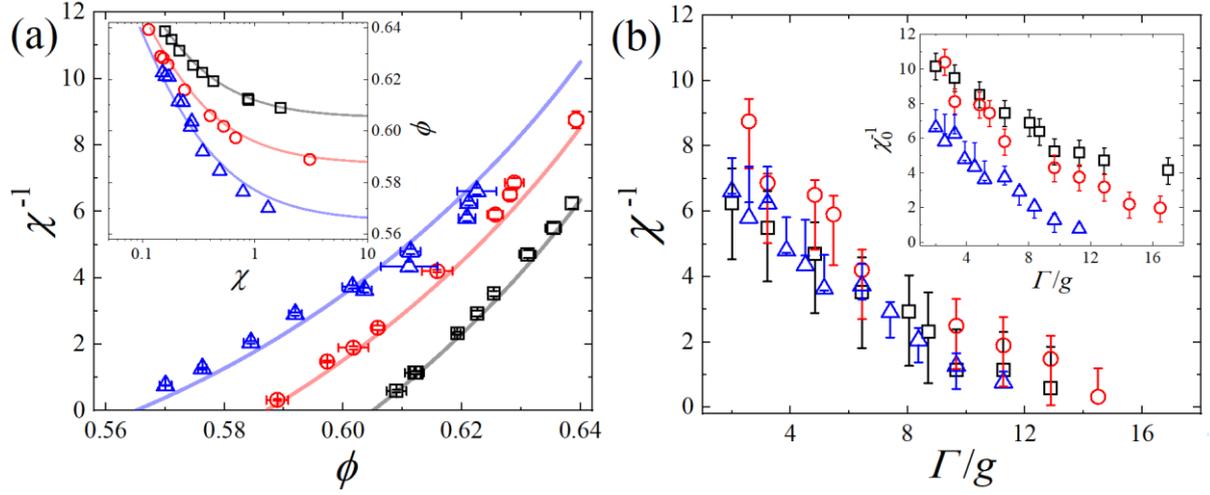

FIG. 2. (a) Inverse of compactivity $\chi^{-1}$ as a function of $\phi$ for the three systems calculated via the overlapping histogram method (symbols) and the fluctuation relation method (solid curves). Inset: $\phi$ as a function of $\chi$. (b) Inverse of compactivity $\chi^{-1}$ as a function of tap intensity $\Gamma$. Inset: $\chi_0^{-1}$ as a function of $\Gamma$, where $\chi_0$ is compactivity calculated by the constant RLP state with $\phi^r = 0.565$. Error bars are evaluated by recalculating $\chi^{-1}$ using Eq. (3) with $\phi^r = \phi^* \pm 0.01$.

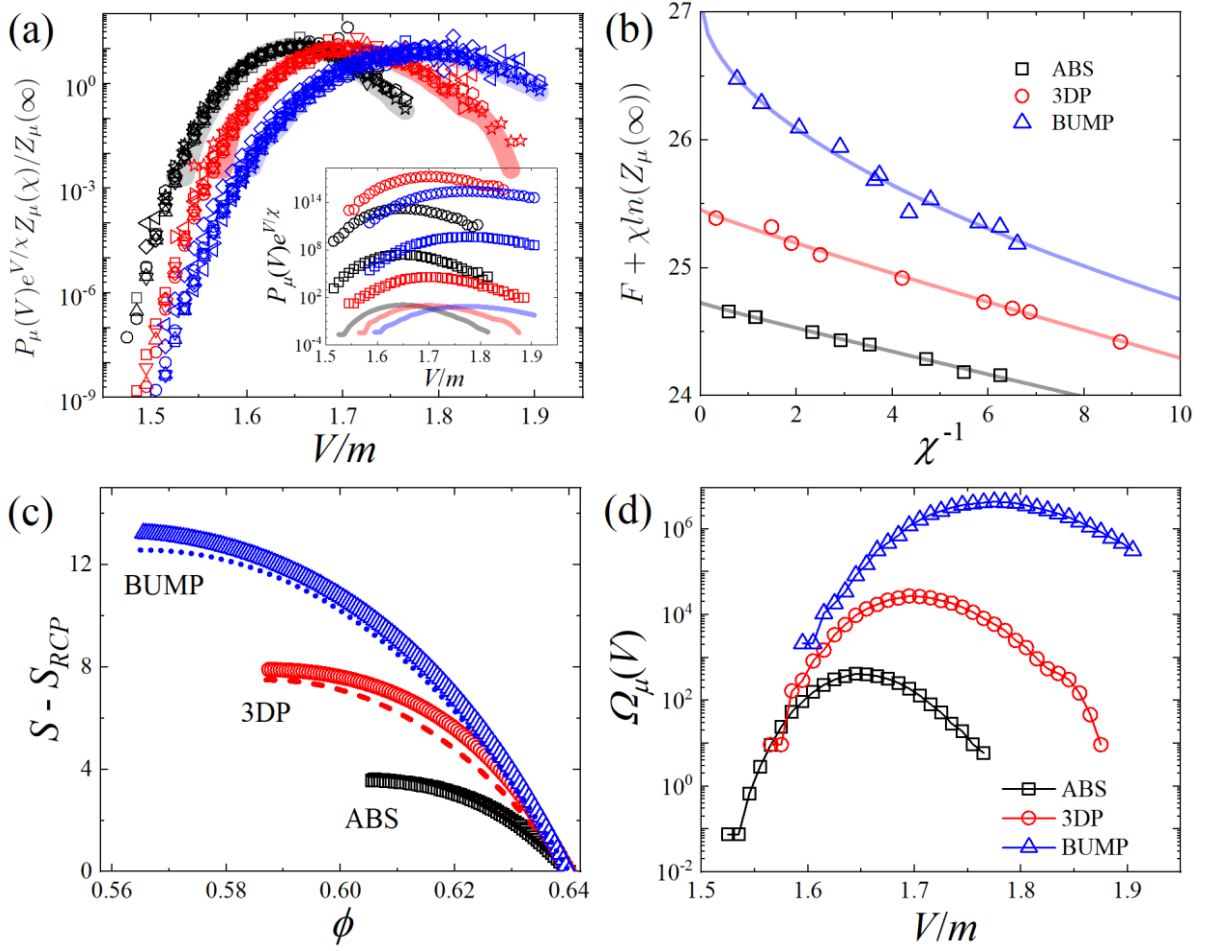

FIG. 3. (a) $P_\mu(V)e^{V/\chi}Z_\mu(\chi)/Z_\mu(\infty)$ as a function of $V/m$ at different $\chi$ (different symbols) collapse on $P_\mu^{\chi\to\infty}(V)$ (thick solid curves) for the three systems. Inset: $P_\mu(V)e^{V/\chi}$ (symbols) at several $\chi$ for the three systems. Solid curves denote $P_\mu^{\chi\to\infty}(V)$. (b) $F + \chi\ln(Z_\mu(\infty))$ as a function of $\chi^{-1}$ by experiments (symbols) and fittings (solid curves). (c) $S - S_{RCP}$ as a function of $\phi$ calculated via $S = -\partial F/\partial\chi$ (symbols) and the fluctuation relation method (Eq. (4), dashed curves) respectively. (d) Density of states $\Omega_\mu(V)$ as a function of $V/m$.

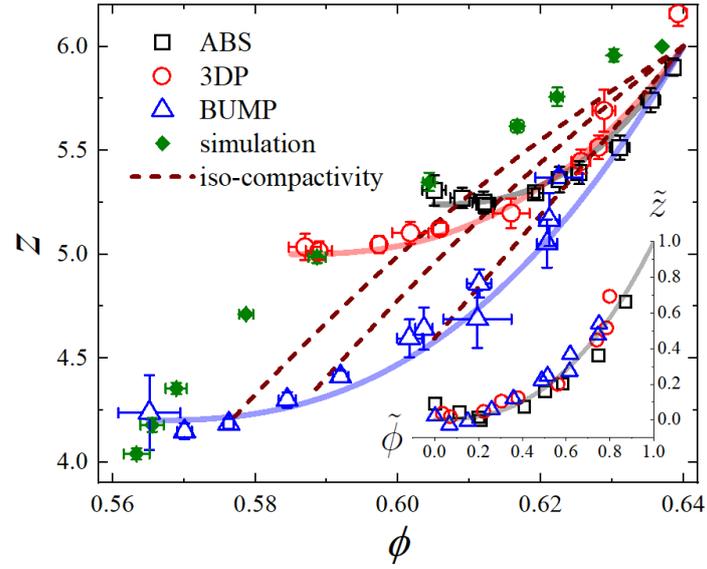

FIG. 4. Granular jamming phase diagram of coordination number $z$ versus packing fraction $\phi$ by experimental measurements (open symbols) and simulation (filled diamonds). Dashed curves denote the iso-compactivity lines with $\chi = 0.8$, $0.4$, and $0.2$ respectively. Inset: the relation between $\tilde{z}$ and $\tilde{\phi}$ (see main text) with an empirical fit $\tilde{z} = \tilde{\phi}^{2.5}$ (solid curve).

# Supplemental Materials

## 1. Bumpy bead

A bumpy sphere model is designed by decorating a given number of small spheres, i.e., bumps, uniformly on the surface of a central large sphere (The centroids of small spheres are on the surface of the large particle). Specifically, we choose the bumpy number as 100 and the diameter ratio between bumps and central sphere as 0.1, as shown in Fig. S1. We manufacture such beads with the circumscribed sphere diameter 5 mm using 3D-printing (ProJet MJP 2500 Plus, 0.032 mm resolution). The bumpy particles can effectively increase interparticle friction by interlocks between bumps when particles contact each other. Since the diameters of the bumps are small, bumpy bead behaves rather similar to sphere as manifested by its reconstructed X-ray tomography images (shown in Fig. S1(b)) and geometrical packing properties.

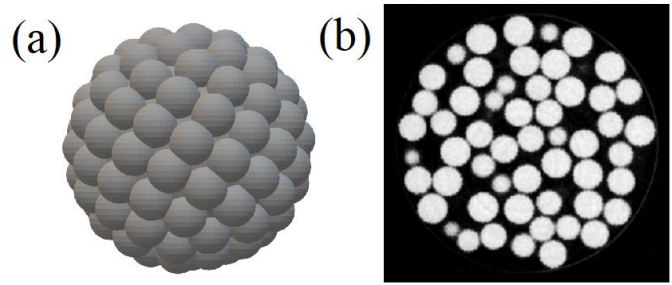

Fig. S1. (a) Visualization of a bumpy bead model. (b) Cross section of a reconstructed packing structure of bumpy beads from X-ray tomography.

## 2. Repose angle measurement

We measure the effective friction coefficients for ABS, 3DP, and BUMP beads by the repose angle method. Beads are fed into a rotating drum with inner diameter 12 cm and height 6 cm, occupying nearly half of its inner volume. Under continuous quasistatic rotation, the system displays increasing unsteady slopes accompanied by avalanches of different magnitudes. We record this process by video imaging and measure the repose angle $\theta$ just before each avalanche. Thus, we can obtain $\mu = \tan\theta$ by averaging over 30 independent realizations with $\mu_{ABS} = 0.609$, $\mu_{3DP} = 0.671$, $\mu_{BUMP} = 0.809$. Note that this value is not equivalent to the microscopic interparticle friction coefficient. But, different definitions for $\mu$ yield the same trend $\mu_{ABS} < \mu_{3DP} < \mu_{BUMP}$.

### 3. Hopper deposition protocol

In the tapping experiment, we approximate $\phi^*$ of a certain bead by the $\phi$ with sufficiently strong tap intensity. To prove the validity of this approximation, here we apply an alternative method, the hopper deposition method, to prepare the loosest packings [5]. A cylindrical plastic tube is vertically placed at the center of the cylindrical container (same one used in the tapping experiment) with its bottom touching the container base and filled with beads. Then, the tube is slowly lifted, discharging the beads gradually. By using different tubes with different diameters, we obtain loose packings with conical top. In general, the larger tube diameter generates looser packings, with the loosest packing fractions $\phi_d = 0.595$, 0.587, and 0.570 for ABS, 3DP, and BUMP, respectively. Note that the associated loosest packing fractions from tapping are $\phi^* = 0.605$, 0.587, and 0.565. Thus, $\phi_d \approx \phi^*$ for all three beads with the error margin about 0.01, indicating $\phi^*$ indeed marks the onset of mechanical stability for packings of a specific bead.

### 4. Coarse-grain of Voronoi volume

According to the original Edwards volume ensemble, Eq. (1) in the main text refers to the total volume of a packing. As discussed in several studies [16, 17, 21], subsystems containing a certain number of particles (denoted by $m$) can be applied to this formulism and the resulting compactivity becomes intensive, i.e., independent of $m$, given $m$ is sufficiently large. A practical method to confirm this point is to calculate $var(V) = \sigma_V^2/m$, where $\sigma_V^2$ is the variance of the total Voronoi volume $V$ with a size of $m$ particles. If $var(V)$ is independent of $m$, the derived compactivity will be intensive, as indicated by Eq. (3). In Fig. S2(a), we find $var(V)$ increases with $m$ and the results for $m = 15$ are approximately the same as those for $m = 30$. Thus, we choose the coarse-grain size as $m = 15$.

The probability distribution function of $V$ will also vary with $m$. In Fig. S2(b), we show $P(V)$ with four different $m$ for ABS packings at $\phi = 0.605$. It is clearly observed that the asymmetric $P(V)$ (typically described by a $k$-Gamma distribution [16]) gradually becomes Gaussian-like and the width shrinks with increasing $m$. In semi-log plots of Figs. S2(c)(d), we find that $P(V)$ still remains asymmetric with the exponential tails when $m < 15$ and thus $m = 15$ is suitable for the overlapping histogram method.

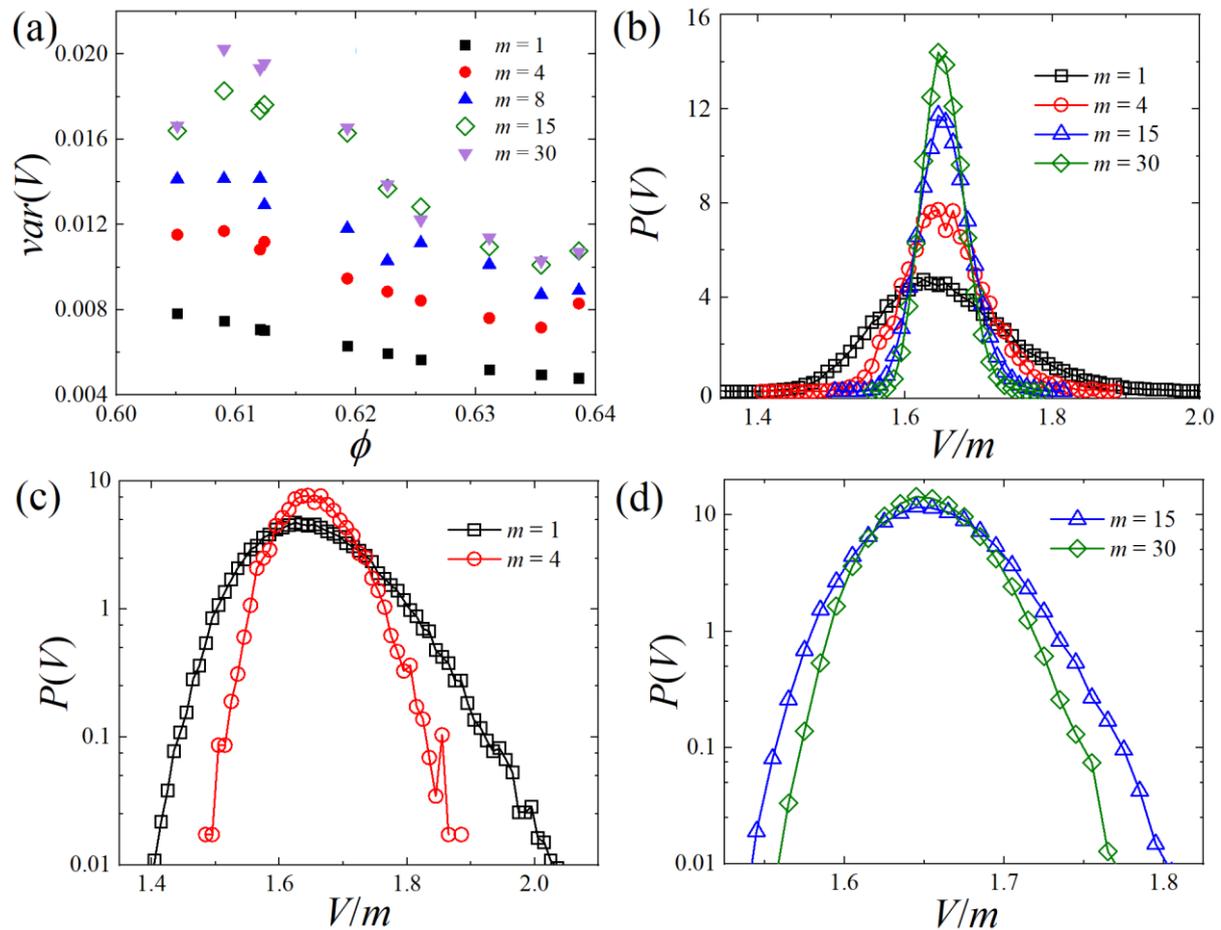

Fig. S2. (a) Relationship between $var(V)$ and $\phi$ with different coarse-grain size $m$ for ABS packings. (b) $P(V)$ with different $m$ for ABS packings at $\phi = 0.605$. (c), (d) Semi-log plots of $P(V)$ for $m = 1, 4$ and $m = 15, 30$ respectively.

## 5. Simulation of frictional jammed packings

We use LAMMPS to perform molecule dynamics (MD) simulations with the spring-dashpot model and static friction (with friction coefficient $\mu$) to generate disordered jammed packings of spherical particles without gravity. The simulation procedure is a quasistatic compression from dilute configurations towards jamming onset, i.e., the obtained packings can well simulate the jammed states of frictional hard spheres with the lowest packing fractions [36].

The model is briefly explained as follows. The potential energy for overlapping particles $i$ and $j$ is given by a linear spring $U_{ij} = et^2/2$, where $e$ is the energy scale and $t$ is the dimensionless overlap ratio (in unit of particle diameter). We randomly initialize an overlap-free configuration of $N = 10000$ particles at $\phi = 0.1$ in a periodic cubic cell. The system is isotropically compressed, i.e., both the boundary length and all the particle centroid coordinates are multiplied by a factor less than 1, with the pre-set packing fraction increment 0.001, followed by MD runs. During the MD process, we frequently check the total potential energy $U = \sum_{i<j} U_{ij}$ and the total kinetic energy $K$. If $U/N < 10^{-10}e$, the system is unjammed and the isotropic compression is repeated. Otherwise, $K$ will vanish with a finite $U$ at a sufficiently large MD step. If $K < 10^{-4}U$, the mechanical equilibrium is approached, generating a jammed packing slightly above the exact jamming onset, i.e., hard-sphere limit. For a simulated jammed packing, both $\phi$ and $z$ are well defined, as presented in Fig. 4 in the main text.

## 6. Influence of friction on local structures

In Fig. 1(d) in the main text, we observe that the volume fluctuation is approximately a function of $\phi$, irrespective of particle types. However, different $\mu$ will lead to the distinctive relation $z(\phi)$, demonstrated in Fig. 4. We find that the statistics of the Voronoi cell volume are robust against specific $\mu$, but the cell itself possesses different structures. In Fig. S3(a), the average Voronoi neighbor number, i.e., the average facet number for Voronoi polyhedra, are clearly different for different beads. Moreover, the difference in structures of Vonoroi cell is also reflected in $N$-ring structures, shown by different fractions of 5-ring structures in Fig. S3(b) [27, 32]. Note that Figs. S3(a)(b) demonstrate a similar fusiform shape as Fig. 4 about $z$, indicating the robustness of both RCP and RLP as the two endpoints.

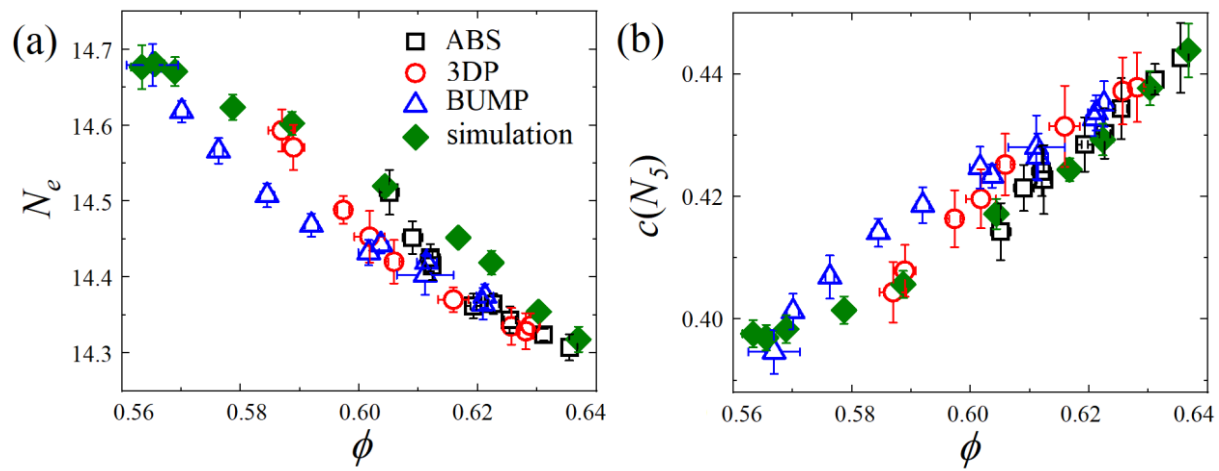

Fig. S3. (a) The average Voronoi neighbor number $N_e$ and (b) the fraction of 5-ring structures as functions of $\phi$ for all three systems, in comparison with the simulated jammed frictional packings.